\documentclass{article}
\usepackage{natbib}
\usepackage{graphicx}
\begin{document}
\newcommand{\beq}{\begin{equation}}
\newcommand{\eeq}{\end{equation}}
\newcommand{\mnras}{MNRAS}
\newcommand{\apj}{ApJ}
\newcommand{\apjl}{ApJ}
\newcommand{\aap}{A\&A}
\newcommand{\araa}{ARA\&A}

\title{
From Molecular Clouds to the IMF: Spatial and Temporal Effects
}

\author{
Shantanu Basu and Sayantan Auddy\\
\\
Department of Physics and Astronomy\\
The University of Western Ontario\\
London, Ontario N6A 3K7, Canada\\ 
basu@uwo.ca, sauddy3@uwo.ca
          }



\maketitle

\begin{abstract}
We review star formation in molecular clouds and describe why magnetic fields may be important and 
how they can influence filamentary structure and the column density probability distribution function (PDF). We also comment on the origin of the stellar and substellar initial mass function (IMF), which may require explanations beyond a simple Jeans length argument in turbulent molecular clouds. A mathematical model of the modified lognormal power-law (MLP) distribution function provides a framework within which to connect accretion processes with the IMF.  
\end{abstract}


\section{Introduction}

Some of the key questions in molecular cloud and star formation research center around the role of magnetic fields. They are invoked as drivers of protostellar outflows, a means of efficiently spreading turbulent and outflow-driven energy through a cloud, and a source of cloud support against gravity. All of the these effects can contribute to the observed low star formation efficiency ($\sim 1\%$) of molecular clouds \citep[e.g.,][]{gol08}. The observed filamentary and wispy structure of many molecular clouds imply that they have an imprint of turbulence as an initial condition. Strong magnetic fields can also help maintain nonthermal motions by driving oscillatory motions in low density regions of clouds and also moderate infall motions to subsonic values in prestellar cores, as is commonly observed \citep{lee99,cas02}. Here, we explore how some observable structural properties of clouds are influenced by magnetic fields, and also discuss the IMF, specifically a model of stellar masses being a result of a killed accretion process.




%
\begin{figure}[]
\resizebox{\hsize}{!}{\includegraphics[clip=true]{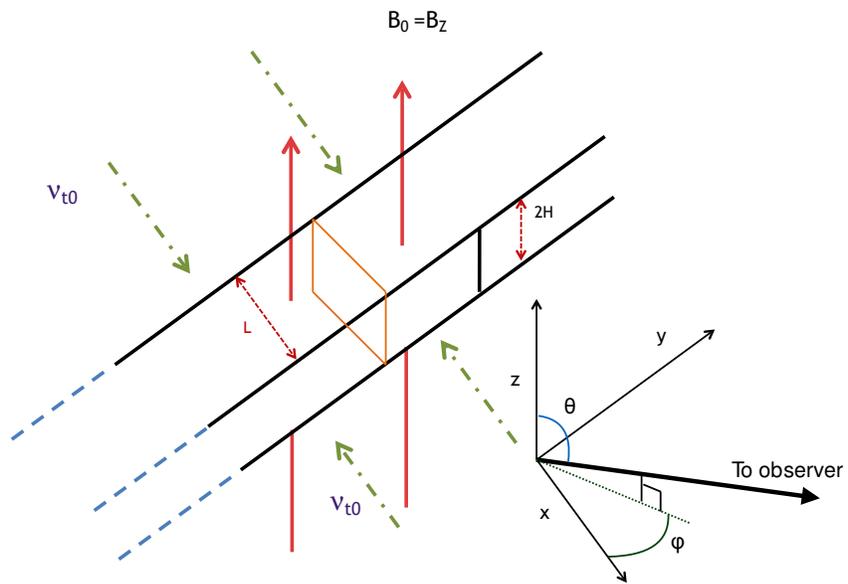}}
\caption{\footnotesize The formation of a magnetic ribbon as the molecular cloud contracts under the influence of turbulent ram pressure and the perpendicular magnetic field. The thick black arrow points to a viewing angle $\theta$.
}
\label{schematic}
\end{figure}

\section{Molecular Cloud Fragmentation}

Molecular clouds can fragment into locally collapsing objects when local gravity dominates over local support mechanisms. The fragmentation can be initiated by a turbulent flow field associated with the formation of the molecular cloud, but fragmentation will also occur in the absence of turbulence. The cause is gravity, either through direct gravitational instability on a dynamical timescale if the cloud has a supercritical mass-to-flux ratio, or due to gravitationally-driven ambipolar diffusion that acts on a longer timescale, if the cloud is subcritical. If the cloud is both subcritical and initially turbulent, an oscillatory and filamentary network emerges, in which isolated regions eventually become supercritical and go into dynamical collapse. These distinct outcomes are presented and discussed in a variety of papers \citep[e.g.,][]{bas04,li04,nak05,kud08,bas09a,bas09b,kud11}.  

\begin{figure}[]
	\resizebox{\hsize}{!}{\includegraphics[clip=true]{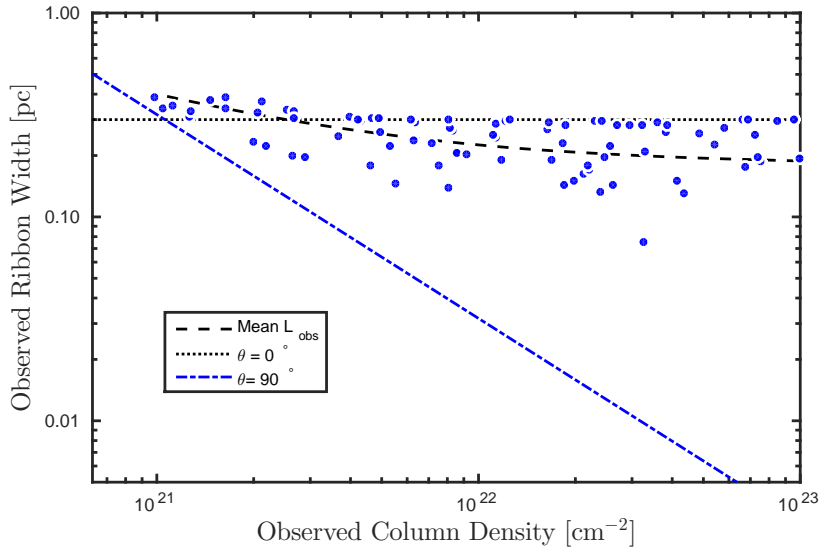}}
\caption{
\footnotesize
Apparent ribbon width $L_{\rm obs}$ versus observed column density $N_{\rm obs}$. Each blue dot corresponds to a magnetic ribbon with intrinsic column density $N$ and observing angle $\theta$. The black dashed line is the mean ribbon width for the entire range of values of $N_{\rm obs}$. The black dotted line is the width when the ribbon is viewed at $\theta = 0 ^{\circ}$. The blue dot-dashed line is the width for the side on view i.e., $\theta = 90 ^{\circ}$.
}
\label{widths}
\end{figure}

The filamentary network of gas column density observed by the {\it Herschel Space Observatory} \citep[e.g.,][]{and10,men10} implies that turbulence may be inherited as an initial condition in molecular clouds. The origin of this turbulence may be instabilities in a post-shock layer \citep{koy02} as the molecular cloud is being formed through large-scale shocks. Since the ambient HI gas of the interstellar medium is largely subcritical \citep{hei05}, it is natural to assume that much of the assembled gas that undergoes conversion to molecules is also subcritical at early times. In \citet{aud16} we showed that filamentary compressions in a subcritical molecular cloud will produce quasi-equilibrium ``ribbons'' in which the magnetic pressure balances the ram pressure of the large-scale flow. Fig.~\ref{schematic} shows a schematic picture of the scenario, in which the ribbon is flattened along the direction of the mean magnetic field and the turbulent compression is primarily in the plane perpendicular to the magnetic field. The lateral thickness $L$ of the ribbon is determined by the balance between ram pressure and magnetic pressure, while the vertical thickness $H$ is set by a balance between internal pressure and gravity. \citet{aud16} show that the lateral thickness is 
\beq
L = L_0 \left[ 2 \left( \frac{v_{t0}}{v_{A0}} \right)^2 + 1 \right]^{-1}, 
\eeq
where $L_0$ is the initial scale of turbulent compression, $v_{t0}$ is the turbulent flow speed, and
$v_{A0}$ is the ambient Alfv\'en speed. If the turbulence is trans-Alfv\'enic, then 
$L \simeq L_0/3$,
which is independent of the local density, unlike the Jeans length, and depends only on the turbulent compression scale $L_0$. This configuration is assumed to have a vertical thickness $H$ that is essentially the Jeans scale, although vertical pressure due to turbulence \citep{kud03,kud06} will also contribute to the thickness in 
general. An interesting quantity to determine is the distribution of observed apparent width of a ribbon configuration when viewed from a variety of viewing angles. Fig.~\ref{widths} shows the result of taking a sample of 100 random viewing angles and plotting the apparent ribbon width $L_{\rm obs}$ against the apparent integrated column density $N_{\rm obs}$. Each blue dot corresponds to one synthetic observation at a random viewing angle. The black dotted line corresponds to the face-on view ($\theta=0^{\circ}$) for which the apparent width equals the lateral width $L$, chosen to be 0.3 pc. The blue dot-dashed line corresponds to $\theta=90^{\circ}$, for which the apparent width is essentially the Jeans length and has a strong dependence on column density. The black dashed line is an average obtained by taking 100 randomly chosen values of $\theta$ for each value of column density and calculating the mean values of $L_{\rm obs}$ and $N_{\rm obs}$. The line is relatively flat over the column
density range $10^{21}$ cm$^{-2}$ to $10^{23}$ cm$^{-2}$, similar to the relatively flat relation observed by \citet{arz11}. 

Another interesting structural property of molecular clouds is the column density probability distribution function (PDF). Observers and theorists alike find this to be an easily measurable quantity, e.g., from dust emission observations and from simulations. Whereas early theoretical work \citep{vaz94} showed that the PDFs are lognormal for turbulent non-self-gravitating media, more recent simulations with self-gravity show that a power-law tail in the PDF is associated with the presence of local regions of gravitational collapse \citep{bal11,kri11,fed13,war14}. Observations also reveal the presence of power laws, and indeed \citet{alv17} (see also \citet{lom15}) claim that there is really no evidence for a lognormal PDF in their sample of clouds, and instead just a power-law profile over a large dynamic range, with a peak in the PDF, if present, set by the presence of a cloud boundary. Alternatively, \citet{pok16} observe the Mon R2 giant molecular cloud and find a lognormal PDF with a power-law extension at high column density. An important question is whether the putative lognormal or power-law PDFs are expected to be universal or depend upon initial conditions. Magnetic fields in molecular clouds are rarely directly measurable through the Zeeman effect \citep{cru12}, so a key goal is also to find structural properties in molecular clouds that can be linked to the ambient magnetic field strength (or mass-to-flux ratio). 

\begin{figure}[]
\resizebox{\hsize}{!}{\includegraphics[clip=true]{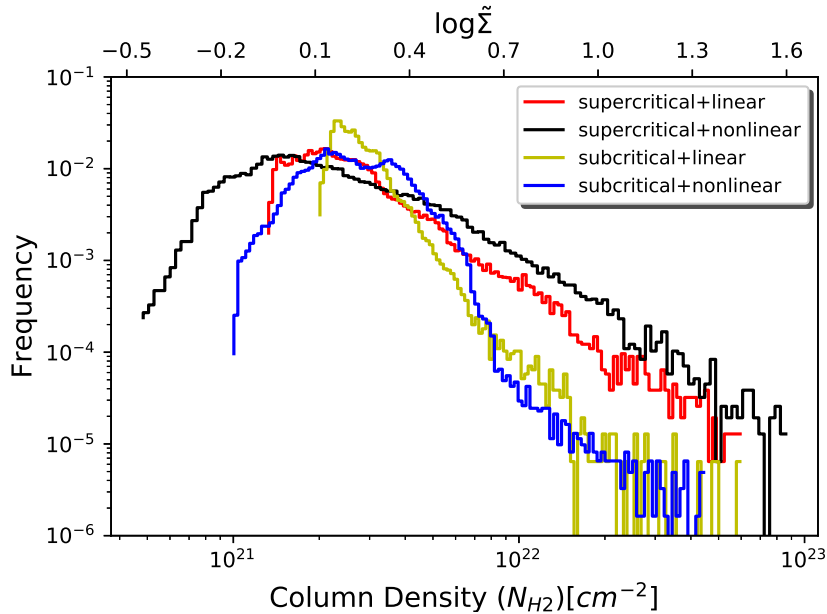}}
\caption{
\footnotesize
The column density PDFs for different initial conditions. The red and the black histograms are supercritical models with linear and nonlinear perturbations, respectively. The yellow and the blue histograms are subcritical clouds with linear and nonlinear perturbations, respectively. The vertical axis is the normalized frequency. Supercritical models have largely a power-law PDF with index $\alpha \approx 2$, a subcritical model with linear perturbations has $\alpha \approx 4$, and only the subcritical turbulent model shows a lognormal body with power-law extension at high densities.
}
\label{comparison}
\end{figure}

In \citet{aud17} we have performed a systematic study of the formation of the power-law PDF due to self-gravity in both non-turbulent and decaying turbulence environments, with either supercritical or subcritical mass-to-flux ratio.   
Key results of this study are in Fig.~\ref{comparison} and also summarized here. For supercritical clouds (weak magnetic field), there is indeed no lognormal PDF. Although the PDF has a peak that is associated with the background column density, there is a steady development of a power-law profile toward higher column densities. This occurs regardless of whether initial conditions are turbulent. The power-law slope is approximately $-2$ when the PDF is binned logarithmically. This is consistent with the development of collapsing cores that have a column density profile $\Sigma \propto r^{-1}$. This PDF can be accurately fit with the modified lognormal power-law (MLP) distribution \citep{bas15}. The MLP is a pure power law in one limit and a pure lognormal in another, depending on the values of its three parameters. The best fit to the supercritical models is a peaked function that has an extended power-law body $dN/d\log \Sigma \propto \Sigma^{-\alpha}$ with index $\alpha \approx 2$. For
a subcritical model with initially small amplitude perturbations, the gravitational contraction is driven by ambipolar diffusion. In this case, there is a slow transition from ambient magnetically dominated regions into a gravitationally collapsing supercritical inner core. Most of the gas is in a transition zone in which the column density profile is significantly shallower than $\Sigma \propto r^{-1}$ and in fact closer to $\Sigma \propto r^{-0.5}$. This effect has also been seen in axisymmetric collapse calculations \citep[see][]{bas95,bas97}. The result is a column density PDF with $\alpha \approx 4$. In the case of a subcritical cloud with nonlinear turbulent initial conditions, the result is the formation of filamentary structures as also described analytically by \citet{aud16}. These structures generally oscillate until ambipolar diffusion creates a supercritical central region that undergoes runaway collapse. The oscillating regions maintain a nearly lognormal PDF while the innermost dense collapsing regions create a power-law tail in the PDF that starts at higher values of $\Sigma$ than in the other models. This creates a PDF that appears to have a lognormal body as well as a distinct power-law tail.

\section{Initial Mass Function}

\begin{figure}[]
\resizebox{\hsize}{!}{\includegraphics[clip=true]{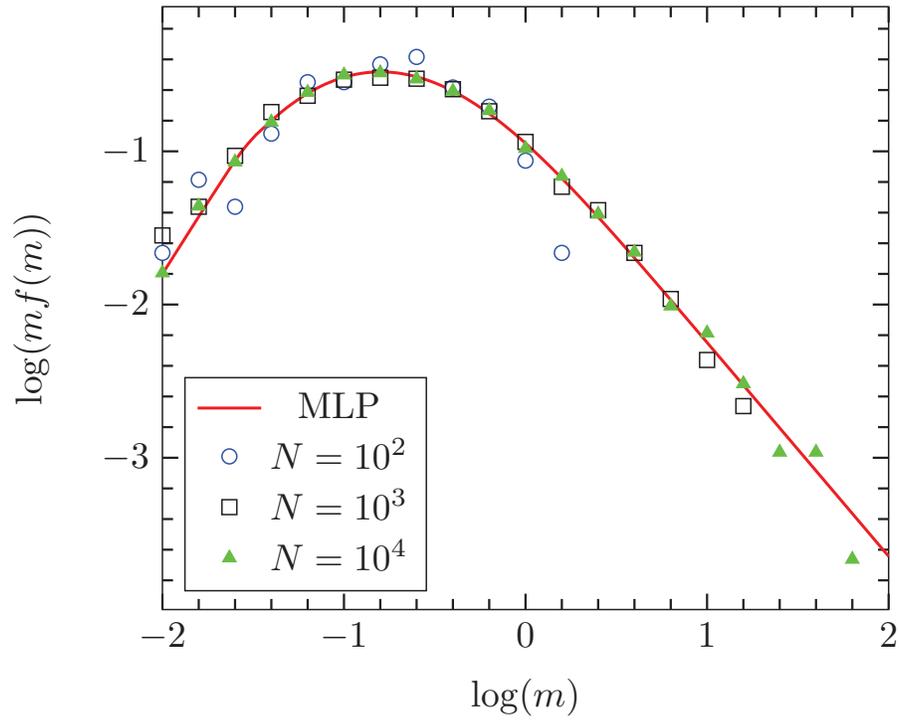}}
\caption{\footnotesize 
	The MLP function with parameters $\mu_0=-2.404$, $\sigma_0=1.044$, and $\alpha=1.396$ (best fit to 
	\citet{cha05} IMF), overlaid with
histogram values for random samples drawn from the distribution
with different sizes $N$ as labeled. All
histograms are binned in increments $\Delta \log\, m=0.2$, where $m$ is in units of $M_{\odot}$. 
The analytic function is plotted as $m\,f(m)$ where $f(m)$ is the density function, and 
the histogram is the fractional number $\Delta N/\Delta \ln\,m$.
}
\label{mlp}
\end{figure}

The structural properties of molecular clouds are often linked to the IMF, through
ideas like a direct conversion of the core mass function (CMF) into an IMF \citep{mot98}.
However, the Jeans mass in dense regions of molecular clouds is typically at least a few $M_{\odot}$,
so that significant inefficiency is needed to account for the large number of very low mass
($\leq  0.1\, M_{\odot}$) stars in the IMF. Furthermore, the discovery of increasing number of brown dwarfs and even free-floating
planetary mass objects is straining this simple picture of a direct mapping from the CMF to IMF.
Both \citet{dra16} and \citet{muz17} find significant numbers of brown dwarfs in young stellar
clusters, and the estimated fraction of substellar objects is approaching that of stellar objects.
\citet{thi15} claim that the theoretical models of a turbulence induced CMF as a root cause of the 
IMF \citep{pad02,hen08} underestimate the observed number of substellar objects, and they posit that
other mechanisms like disk fragmentation are needed to explain the low mass range. Indeed the 
simulations of self-consistent disk formation and evolution show that ejections of proto-substellar 
objects may be quite common in the early phase of disk evolution \citep{bas12,vor16}.
Whereas filamentary structure and PDFs may inform us a lot about the levels of turbulence and 
magnetic fields in molecular clouds, it seems that the origin of stellar masses may rely on the 
concept of star formation as a killed accretion process. 
See \citet{ada96} for an earlier version of an accretion 
termination model for the IMF. In this view, stellar seeds start with masses as low
as $\approx 10^{-3}\, M_{\odot}$, the mass of a second stellar core, or $\approx 10^{-2}\, M_{\odot}$, the mass of a first hydrostatic core, and accrete from their surrounding
core or disk until some event terminates their accretion lifetime. This could be caused by ejection
from a multi-body system, by outflows from the object or a companion sweeping away matter, or
by a very disruptive event like feedback from a high-mass star in the cluster. If we assume an equally likely stopping of accretion in each time interval, then the lifetimes follow an exponential distribution $f(t) = \delta\, e^{-\delta t}$. If the mass of each seed grows according to $dm/dt = \gamma m$, then the resulting normalized pdf for masses after accretion termination is
\begin{eqnarray}
	f(m)  & =  & \frac{\alpha}{2} \exp \left[ \alpha \mu_0 + \alpha^2\sigma_0^2/2 \right] \: m^{-(1 + \alpha)}  \nonumber \\ 
	& & \times \: {\rm erfc} \left[ \frac{1}{\sqrt{2}} \left( \alpha \sigma_0 - \frac{\ln m -\mu_0}{\sigma_0} \right) \right].  
	\label{mlpeqn}
\end{eqnarray}
This is the MLP distribution, in which $\mu_0$ and $\sigma_0$ are the mean and dispersion of an initially lognormal distribution of masses which then undergo 
accretion growth, and $\alpha = \delta/\gamma$ is the dimensionless ratio of ``death'' rate to ``growth'' rate of protostars. The exponential growth of mass may be relevant for the formation of intermediate and high mass stars because of the relatively small age spread of young stars of widely different masses \citep{mye93}. This implies that the accretion rate needs to increase rapidly in order to form higher mass stars. The initial mass accretion rate may however be more like a constant \citep{shu77} and a hybrid model with an initially constant mass accretion rate that later transitions to exponential growth, coupled with an exponential distribution of accretion lifetimes, has been developed by \citet{mye14}. 

A fit of Eq.~\ref{mlpeqn} to the IMF of \citet{cha05} is shown in Fig.~\ref{mlp}. This fit with the MLP function \citep{bas15} requires only three parameters, whereas the approach of manually joining a lognormal with a power law (as adopted by \citet{cha05}) requires four parameters (two parameters of the lognormal, a joining point, and a power-law index). According to this fit, 24\% of objects are substellar ($< 0.075\, M_{\odot}$), accounting for 2\% of the total mass. The MLP model 
hypothesizes that the initial distribution of accreting protostellar seed masses is lognormal. However,
a lognormal of very small dispersion $\sigma$ approaches a delta function, and we may also think of the 
initial distribution as being a very narrow (delta-like) function at the mass of the first hydrostatic cores, $\approx 10^{-2}\, M_{\odot}$ or even of the second stellar cores, $\approx 10^{-3}\, M_{\odot}$. An early termination of accretion for many seeds can naturally account for a large fraction of objects being substellar or very low mass stars.


\section{Summary}

We have presented a scenario for turbulent molecular clouds with subcritical mass-to-flux ratio in which oscillatory quasi-equilibrium ribbons can be formed and have apparent widths that are largely independent of column density.
Furthermore, we have summarized a study of the systematic dependence of the column density PDF of molecular clouds on magnetic field strength and turbulence, with observationally distinguishable outcomes between supercritical and subcritical initial conditions. Most cases develop a direct power law, and only the subcritical clouds with turbulence are able to maintain a lognormal body of the PDF but with a power-law tail at high values. Finally, we have presented a scenario of star formation as a killed accretion process, which may be relevant to understand both the very low mass and high mass ends of the IMF.

\vspace{0.25in}
{\small
SB is sincerely indebted to Francesco Palla for the many insightful discussions about star formation,
for his thoughtful advice, and for attending the very stimulating conferences IMF@50 and SFR@50
that he organized. We also thank our collaborators M. Gil, T. Kudoh, and E. Vorobyov.
}

\end{document}